# Restoring Vision through Retinal Implants - A Systematic Literature Review

**Magali Andreia Rossi [1], Sylviane da Silva Vitor [2]**

[1]São Paulo State Technological College, Fatec Carapicuíba/Fatec Osasco
[2]Itau Unibanco S/A

[1]mandreiarossi@gmail.com, [2]sylvianevitor@gmail.com

*Abstract*–This work presents a bunched of promising technologies to treat blind people: the bionic eyes. The strategy is to combine a retina implant with software capable to interpret the information received. Along this line of thinking, projects as Retinal Prosthetic Strategy with the Capacity to Restore Normal Vision from Weill Medical College of Cornell University Project, Update on Retinal Prosthetic Research from The Boston Retinal Implant Project, and Restoration of Vision Using Wireless Cortical Implants from Monash Vision Group Project, have shown in different context the use of technologies that commits to bring the vision through its use.

## 1. INTRODUCTION

The possibility to develop visual perception by electronically stimulating the brain has been studied since 1950, when Wilder Penfield observed there was a rational map of visual space onto the primary visual cortex (PENFIELD; RASMUSSEN, 1950). Looking beyond, cortical vision prosthesis date from 1999, when scientists from University of Utah developed a penetrating electrode array to stimulate the brain and recreate a limited form of functional vision: the phosphenes (NORMANN et al., 1999). Bionic eyes and retinal prosthesis nowadays just extend these concepts adapting materials and developing an outstanding processing power.

This work subsumes strategies and the most recent technologies to restore vision, covering from image processing appliances to retinal and cortical implants. It was focused the analysis on three main projects: "Retinal prosthetic strategy with the capacity to restore normal vision: Weill Medical College of Cornell University Project" (NIRENBERG et al., 2001), "Update on retinal prosthetic research: The Boston Retinal Implant Project" (RIZZO, 2011) and "Restoration of vision using wireless cortical implants: Monash Vision Group Project" (LOWERY, 2013).

This analysis is to comprehend the state of art of retinal implants functioning and construct a background about the technologies developed until nowadays.

This work is structured as follows. Section II presents related projects, relating biological solutions that evolve, in different levels, a hacking prospective. Section III presents an explanation of the research projects in which the proposal is restore the vision through retinal implants and brain waves patterns. Section IV presents comparing results acquired from the detailed analysis of the three main projects. Section V discusses the target presented to long of this paper.

## 2. RELATED PROJECTS

The two following purposes manipulate biological structures to treat a series of degenerative diseases and, particularly, to eye diseases. The first one, Clustered Regularly Interspaced Short Palindromic Repeat/Cas 9 (CRISPR/Cas9) technology is an innovative molecular analysis technique used to edit genomes. The CRISPR consists of short segments of DNA, composed by nucleotides, each one adjacent to a spacer DNA – non-codifying section of DNA – inserted after contact with foreign genomes (SUZUKI et al., 2016). Combining CRIPR-Cas9 with Homology-Independent Targeted Integration (HITI) allowed scientists to scientists improve visual function. They tested the site-specific transgene integration using the Royal College of Surgeons rat model for retinitis pigmentosa. The HITI method establishes new avenues for basic research and targeted gene therapies (YANG et al., 2013) including treatments for conditions that causes blindness to humans.

The second one, uses stem cells to treat eye diseases. Stem cells primary characteristics are their unparalleled regenerative capacity and flexibility to grow into different types of cells. Cell-replacement therapy and ex vivo gene therapy are two strategies that take advantage of these properties to treat degenerative diseases and could be expanded, particularly, to treat eye diseases, such as AMD and retinitis pigmentosa. The idea is to explore neural stem cells-progenitors properties for brain repair or also retinal cells replacement. Although these techniques still depend on gathering information about mechanisms that regulate their proliferation and differentiation; they are promisors and another biohacking perspective to restore vision.

# 3. RESEARCH METHODOLOGY

## 3.1   Systematic Literature Review

The Systematic Literature Review is defined as study to map, to evaluate, identify gaps, deficiencies, reinforce as well as find results of primary studies from research developed or in progress in the area. Second Cochrane handbook, define a systematic review as 'uses explicit, systematic methods that are selected with a view to minimizing bias, thus providing more reliable findings from which conclusions can be drawn and decisions made' (HIGGINS; THOMAS, 2019).

The use of Systematic Literature Review consists in looking for answers from literature review based on relevant studies.

## 3.2   Research Question

The focus of the research question is to verify and produce information relative to rising up proposed technologies in this area. Analyzing medical approaches and engines that explore innovative solutions to restore vision was a long way journey because references are scarce and most of them comprehends only retinal prosthesis, which is not capable to develop normal sight yet. The three projects shown in Section 4, on the other hand, tends to a new perspective about how engineering engaged to medicine can promote encouraging results, in cases that simply pathological treatment has no longer effect.

# 4. COMPENDIUM OF THE AVAILABLE PROJECTS

In this section the projects considered as state-of-art are presented. All selected projects were analyzed to obtain their technical details that verify the efficiency of use by human beings.

## 4.1   Weill Medical College of Cornell University Project: Retinal Prosthetic Strategy with the Capacity to Restore Normal Vision

Retinal prosthetics are the only hope for the vast majority of patients with retinal degenerative diseases. Therefore, improve prosthetics performance is the main object of project developed by scientists from Weill Medical College of Cornell University. They focused on a critical factor of prosthetic capability: the retina's neural code.

This prosthetic tool captures the image transformation that occurs naturally via retinal processing and produces the same patterns of action potentials that retina would normally produce. It consists of two parts: an encoder; which makes transformations based on these performed by retina and a transducer; which guides the ganglion cells to fire as the code specifies. This new gadget can be used to offer diversified design solutions, implementing the code into external devices.

Designing a system composed by the encoder and transducer was only the first step. Several evaluations had been done to configure that this retinal prosthetic really brings the benefits it intends to.

In this case, animal testing was impracticable, once millions of test sets would demand an extremely large number of animals. That is why a standard mouse model of retinal degeneration has been used to simulate blind retinas. To show the results three sets of recordings were used. This device has shown encouraging results, not only quantitatively but also qualitatively. The performance – quantified by the fraction of times the responses were correctly decoded – measured for the blind retinas with the encoder was 96%. Thus, with the standard method was only 25% (NIRENBERG; PANDARINATH, 2012). This effectiveness proves that the encoder component is critical; without it, tracking responses are random.

The most significant part of this project is not only the dramatically increase of prosthetic performance but also a new range of possibilities. First, the possibility to, instead of trying to recreate computations in the retina, to make it in an external device: the encoder-transducer system.

It allows significant upgrades to reach restoration of normal sight. Even though the high hit rate, over 90%, this micro effectiveness is not able to develop normal sight yet, and still creating just a small impression of spotlights. It occurs because even if the code responses are very close to normal ganglion cell; the retina output cells remain been used to transmit the electrical pulses to the brain.

That is, it still demands their normal operation to made prosthetic functional and patterns recognition still not able to be properly transferred to where the real processing occurs: the brain.

### 4.1.1 Design

The duo encoder/transducer may seem uncomplicated at first, but the whole gadget to perform and deliver data as a normal retina require a most complex system and computational resources. The Figure 1 represents the design of the components, in a detailed way.

Figure 1. Technical Link between the Components of Cornell University Project.

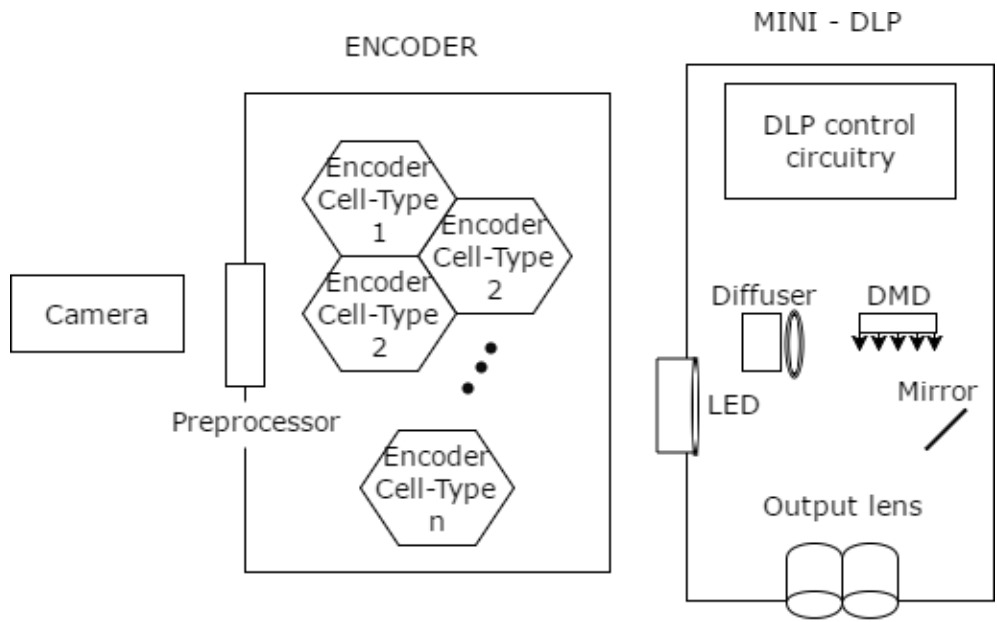

Font: Adapted from (NIRENBERG; PANDARINATH, 2012)

The first component – an infrared video camera. This software (representing the eyes) finds the horizontal and vertical coordinates of the centers of the corneal reflection and the pupil (sampled at 60 Hz) and measures the difference between these two points along the horizontal and vertical axes. The image captured by the camera is delivered to the front end preprocessor, whereby the light level and contrast range are rescaled to fall within the operating range of the encoders. The encoder is a retinal input/output model generated from data taken from the normal retina and also where most of the computational resources are hidden. Images that come and are convolved with a spatiotemporal filter, them are passed through a linear-nonlinear (LN) cascade built to capture stimulus/response relations for a broad range of stimuli; followed by a Poisson generator, that converts them to spikes. In brief, the model captures in a compact form the processing that the retina performs, including center-surround organization, temporal filtering, and spike generation.

The encoder consists of arrays of smaller encoders, one for each ganglion cell type. The different encoders all use the same model structure, the LN cascade, differing each other only by the numerical values of the parameters. However, the number of parameters is too large to resemble all ganglion cell types and, to address this problem; scientists developed a machine learning approach. With a minimal enlargement of the stimulus set (white noise and particular natural scene movies), they obtained a linear-nonlinear cascade model that generalized broadly (white noise, gratings, landscapes, people walking, cars driving, faces, animals, and another one). The pulse patterns resulted from the encoder work are then converted to blue light pulses, which are delivered to the retina using a Mini Digital Light Projector - MINI-DLP, modified to deliver 0.7mW/mm2 of 475 nm light at the retina using a Light-Emitting Diode - LED, showed on the third block of Figure 1. It also contains a digital micromirror device - DMD, which is a grid of mirrors whose positions can be switched with very high spatial and temporal resolution. This switching is controlled by signals from the encoder, allowing light from the LED to reach the retina.

**4.2 The Boston Retinal Implant Project: Update on Retinal Prosthetic Research**

Researchers from Boston Retinal Implant Project focused on creating an advanced device which risks were justifiable. They developed a wireless device implanted in the sub retinal space and created a hermetic system capable to deliver individually electrical stimulation – controlled pulses – to each of hundreds of electrodes. The current device is composed of both internal and external parts. The external part consists of special glasses equipped with a camera – to record the environment and a transmitter – which wirelessly deliver power and the recorded data to the internal part: a photodiode situated behind the pupil, which creates more power and transmits the encoded data to a small decoder chip located on retinal surface that sends an electrical current to individual electrodes to stimulate the retina.

The special glasses are responsible for the "Input", that is, just capture and transmit data. The parts inside the eye are responsible by powering the electrodes, transforming the images captured into a spatial pattern of controlled electrical pulse and delivering these pulses to stimulate the retina (KLAUKE et al., 2011), which will re-process them. In this project, data is differently processed, because instead of reproducing all retinal functions using the photodiode and decoder, the retina still being used, after stimulus, to better deliver data to the brain. The Boston Implant Project developed prototypes using two methods to send data wirelessly: laser light and radiofrequency. There were also two prosthetic designs, differing only in the number of electrodes; to ascertain the hypothesis that more electrodes will provide better vision. The device that enters the eye is made of polyimide

(plastic) because of its flexibility, necessary to bend to the retina's curvature.

For the surgery, they developed a method that requires only a single slit in the back of the eye to introduce a 5-mm-diameter electrode array; the largest surface area of an array safely implanted into a living eye. This package of changes is responsible for distinct results from standards surgical methods, overcoming the spotlights perception.

Studies demonstrated that most of the implanted patients were able to make out spots of light, but also tell the difference between them, this is important, because shows that the patterns are not random anymore. Patients with Retinitis Pigmentosa could localize the quadrant of a large stimulus; identify the direction of moving lines and some common objects like forks and cups too (YANAI et al., 2007). "One of the most encouraging results was one patient who, after one week of implantation, was judged capable to read letters and words similar in size to those in newspaper headlines" (RIZZO, 2011). The BRIP reaches closely 5% of the patients with eye's diseases, almost 2 million people (ALLIANCE FOR AGING RESEARCH, 2012). The complex system involving internal and external parts has several advantages on biocompatibility aspects, but makes the costs even higher. The researchers noted some limitations that challenge the device to restore perfect vision, perhaps the most critical ones are:

1. Find biocompatible material to sell hermetically the device that remains intact when implanted.

2. Create an electrode array that will not decay over time because of cell growth.

3. Identify the best areas of retina to stimulate to achieve the highest quality vision.

Only after these improvements, the device would be able to promote their goal, allow patients to "navigate safely in an unfamiliar environment" (RIZZO, 2011) and to get closer to restore perfect vision.

### 4.2.1 Design

This project concerns different combinations of processing that demand a more meticulous analysis. Understanding the functioning of devices is a valuable way to reach this goal, the Figure 2 will be used for this explanation.

Figure 2. Technical Link between the Components of BRIP.

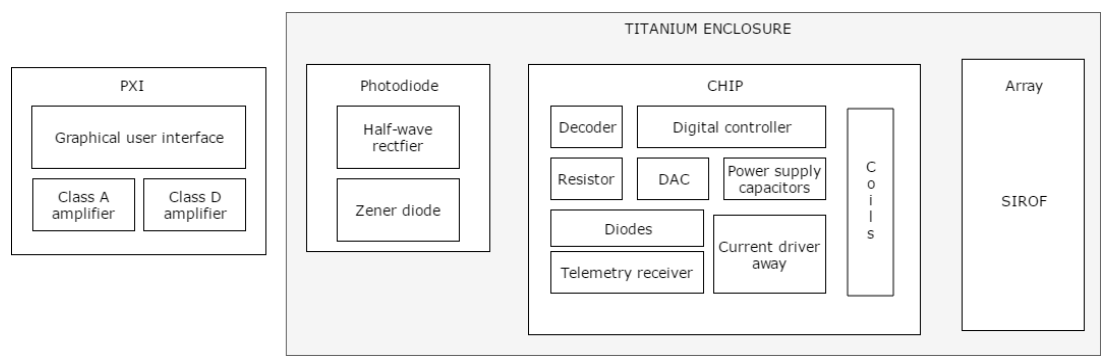

Font: Adapted from (RIZZO, 2011).

The transmitter described in Figure 2 is responsible by delivering data to the implanted part, but this transmission is not random and has to follow several specifications. To determine and configure stimulation pulse strength, duration and frequency, a Peripheral Component Interconnect eXtensions for Instrumentation (PXI) computer was used, the selection is made using the graphical user interface. In the first block of Figure 2, PXI is shown with two different amplifiers that transmit the commands and power from the external computer to the implanted components by near-field inductive coupling.

Class A amplifier is responsible by delivering stimulation data with a 15.5MHz carrier, amplitude shift keyed at a 100% modulation index (SUZUKI et al., 2016). Data is delivered as 16-bit commands and the 170-bit configuration package, these bits are encoded by pulse width modulation, with 30% duty cycle representing a digital bit 1 and 50% representing a digital bit 0. Class D amplifier and a series resonant tank with 125KHz carrier are responsible by delivering power to the dual half-wave rectifier, shown inside the Photodiode block of Figure 2. This power creates a ±2.5V anodal and cathodal supply, which is clamped by the 5.1V Zener diode (KELLY et al., 2011). The custom integrated circuit receives the incoming data beyond the telemetry receiver; decodes them, using the decoder, and delivers stimulating current to the appropriate electrodes based on the timing of transmitted commands, through the primary and secondary coils. The chip can deliver up to 930µA per channel in 30µA steps, using a linear current Digital-to-Analog Converter (DAC) and the current driver away. The other elements, as diodes, resistor and capacitors just guarantee the proper functioning of the chip.

The fourth block of Figure 2 shows the micro fabricated thin-film polyimide array of Sputtered Iridium Oxide Film Electrodes (SIROF) that delivers electrical stimulation current to the retinal nerve cells. Innovation takes place in the serpentine design of the electrodes that allows surgeons to route it under the superior rectus and insert them in the superior-temporal quadrant. All implanted electronics are encased in a titanium enclosure, to sell hermetically the devices and also serve as a current return counter electrode for the stimulation. The computational power of the PXI, combined with innovative strategies to deliver power and data brings BRIP to a new category of retinal devices, but still not the most effective way to restore vision.

## 4.3 The Monash Vision Group Project: Restoration of Vision Using Wireless Cortical Implants

The majority effort on restoration of sight has gone into developing retinal implants to replace damaged photoreceptors (LOWERY et al., 2015). Monash Vision Group researchers are developing another access point to the visual pathway: the Visual Cortex (V1). They created a bionic eye based on implanting small tiles in this region, using penetrating electrodes.

The electronic system is composed by: a high resolution miniature digital camera – which captures the image and send them to a custom – designed "pocket processor", that extract the most useful features from the image; a wireless power and data link, which power and deliver data to the multiple implant tiles that contain, each one, a receiver coil, tuned to the transmitter.

A diode rectifier is used to extract power and its output and to supply the Application Specific Integrated Circuit (ASIC), which contains sufficient registers for each electrode be set individually via the wireless link. The Figure 3 shows how the top components are related.

Figure 3. Functionality Diagram of the Monash Vision Group Project.

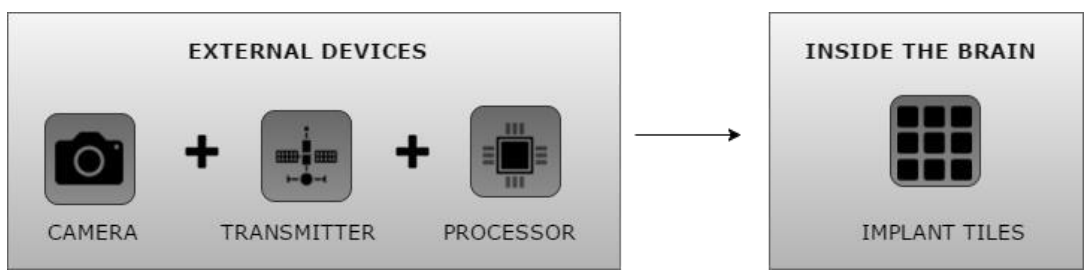

Font: Authors.

External devices in Figure 3 shows four phases of data processing: "Input" – promoted by the camera, "Pre-process" and "Process" – by the pocket processor and also "Output" – by the transmitter. Inside the brain, the second block of Figure 3 is responsible only by delivering the cortex V1 separately each visual sensation, but this flexibility converges to a higher perception owing to an effortless manipulation of data received at "Input" phase. The surgical technique is to implant the electrodes through pia mater – a thin and strong covering of the brain. To treat individual differences between people's brain, visual cortex anatomy will be imaged using high-resolution Magnetic Resonance Imaging (MRI). The method of stimulation is to inject a negative current from the electrodes inserted into the surface of cortex.

This new purpose expects to increase resolution, that is, to overcome the spotlights and really reach normal vision. Human tests have been expected, but it is already known that, for a given spacing of electrodes, it is possible to obtain a 20x advantage in resolution over a retinal implant (LOWERY et al., 2015).

Otherwise, provided stimulation is limited to "on and off", because of the restricted electrode number, and other properties that compose perception such as space-time, density, texture and another one, still not being contemplate.

This project brings an electrical engineering approach to a biomedical engineering problem, providing a defined image directly to the brain. However, this innovation leads to a new challenge: how to keep visual acuity with a 2D image that, even when processed and provided by a high-definition camera, does not improve resolution?

**4.3.1 Design**

Signal processing computations of images to provide a coded output destined to the brain demands a huge computational power, which lodges inside the Pocket Processor, this portable computer, detailed in Figure 4, contains his own operating system and drivers for the off-the-shelf digital video camera, wireless link and programming interfaces.

This package also includes a high capacity, lightweight battery that powers the camera, the processor itself and the implant electronics; it can power multiple tiles up to 25mm away and consume less than 400mW to maximize battery life ("Ghana Health Service Annual Report 2014," 2015), (DUA; SINGH, 2010). The battery is charged using the same socket that is used by the wireless link, so that there is no possibility of the user connecting the system to the mains while it is in use.

Controls are, basically, push buttons with braille labels, haptic and audio feedback; allow the user to program the processor to work on three different modes, based on the following algorithms: edge detection – shows the outlines of objects such as tables and door frames; people detection – identifies the presence of people in the user's environment and ICP – localizes themselves within their environment.

Figure 4. Technical Link between the Components of MGV Project.

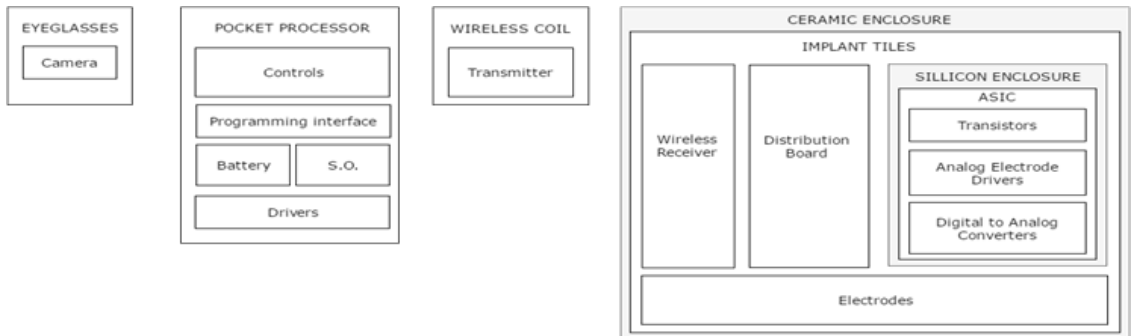

Font: Adapted from (LOWERY et al., 2015).

Intensity Thresholding and Integral Sample algorithms were also used to extract the most important features from the images. The Pocket Processor is connected to a wireless coil, whose transmitter sends the coded output and a clock signal to the implant tiles, defining when each data bit has been sent. An electronic circuit simulation software and a prototype transmitter circuit compose the transmitter, this system converts signal received from the processor to the modulation format: a modified form of pulse-width modulation. The fourth block in Figure 4 shows the implant tiles, composed by a wireless receiver, a distribution board, an Application Specific Integrated Circuit (ASIC) and 45 platinum-iridium electrodes. The first one is directly connected to the transmitter includes a digital logic circuit incorporating approximately 500,000 transistors (AUSTRALIAN RESEARCH COUNCIL, 2011). The distribution board is a sub-assembly interface that provides connectivity between the ASIC, the electrodes and active electronic components. The implant themselves are also programmed to decode data and recognize the appropriate currents to each of their 45 electrodes.

## 5. RESEARCH RESULTS

The three main projects represent an overview of the state of art for treatments to restore sight. To evaluate the variability of these technological contributions was developed the four following tables. Table 1 shows that, although the projects intend to restore vision by purposing a general solution - reestablishing the communication channel - they are limited to blindness caused by degenerative diseases and, in Weill Medical College of Cornell University Project, to even more specific cases. Moreover, the leading cause of blindness cataract cannot be solved by any of these techniques.

Table 1. Applicability projects on treatment for different blindness causes:
degenerative diseases, congenital blindness, visual impairments and cataractblindness.

| APPLICABILITY | Weill Medical Collegeof Cornell University Project | Boston Retinal Implant Project The Monash VisionGroup Project | Monash VisionGroup Project |
|---|---|---|---|
| Degenerative diseases | Yes | Yes | Yes |
| Congenital blindness | No | No | No |
| Visual impairments | No | Yes | Yes |
| Cataract blindness | No | No | No |

Font: Authors.

On computational perspective, the processing algorithms demand special attention, considering that they are responsible by most of the quality of data transmitted, so intrinsically related to the visual acuity delivered.

The Table 2 shows that, in terms of computational power, none of the projects brings up innovative solutions, they use algorithms that, in most cases, does not offer an efficient compression method and even by taking advantage to artificial intelligence, does not provide verified efficacy.

Table 2. Processing algorithms of Weill Medical College projects technical
information: AI application, language level, compression methods and testability.

| PROCESSING ALGORITHMS | Weill Medical Collegeof Cornell University Project | Boston Retinal Implant Project The Monash VisionGroup Project | Monash Vision Group Project |
|---|---|---|---|
| AI application | Yes | No | Yes |
| Language | Low level | Low level | High level/ Low level |
| Compression methods | Low effective | Unknown | Medium effective |
| Testability | Undefined | Undefined | Low |

Font: Authors.

Innovative implants are normally related to outstanding hardware approaches or surgical techniques. With this in mind, Table 3 brings up that, in terms of architecture, two projects still depend on the external devices to properly functioning. In addition to that, their main workflow is the same: external devices capture and processes data and internal devices deliver them.

Table 3. Hardware approaches projects considering external devices, architecture, battery life and environment dependence.

| HARDWARE APPROACHES | Weill Medical College of Cornell University Project | Boston Retinal Implant Project The Monash Vision Group Project | Monash Vision Group Project |
|---|---|---|---|
| External devices | Yes | Yes | Yes |
| Architecture | Externally dependent | Externally dependent | Biocompatible |
| Battery life | Undefined | Unknown | Medium |
| Environment dependence | Low | Medium | Medium |

Font: Authors.

## 6. DISCUSSION

Restoring sight is a challenging goal, but also a global health demand. This paper spread throughout three projects, outlining their transforming power and results. However, an analysis about this progress cost-effectiveness is necessary to evaluate which barriers we still face to reach this goal. In 1998 (NORMANN et al., 1999), scientists already concerned about retinal implants problems, such as: its limitation to outer retinal pathologies; the inefficiency to provide discriminable patterned percepts instead of generalized spots of light and the unguaranteed encoding of spatial, chromatic, intensity and temporal information by device.

Both software and hardware developed by the three main projects has been carefully analyzed by their major aspects (applicability and algorithms), as shown from Table 1 to 3. These evolving technologies reveal a panorama about how close the scientists are to restoring normal vision.

Projects that demonstrate a high level computational power to process the images, without regard to how communication channel – hardware – will work are considered challengers, because demand large attention to the physical and sometimes biological aspects. In the other hand, a complete hardware, with proven efficacy to deliver information, but few integrations with appropriate processing algorithms is considered visionary.

The solution which combines the two aspects, delivering high visual acuity is classified as outstanding. However, most of the projects still in the regular baseline, considering their few innovative and little expressive results, as shown in Figure 5.

Figure 5. Magic Quadrant of Projects Innovative Level.

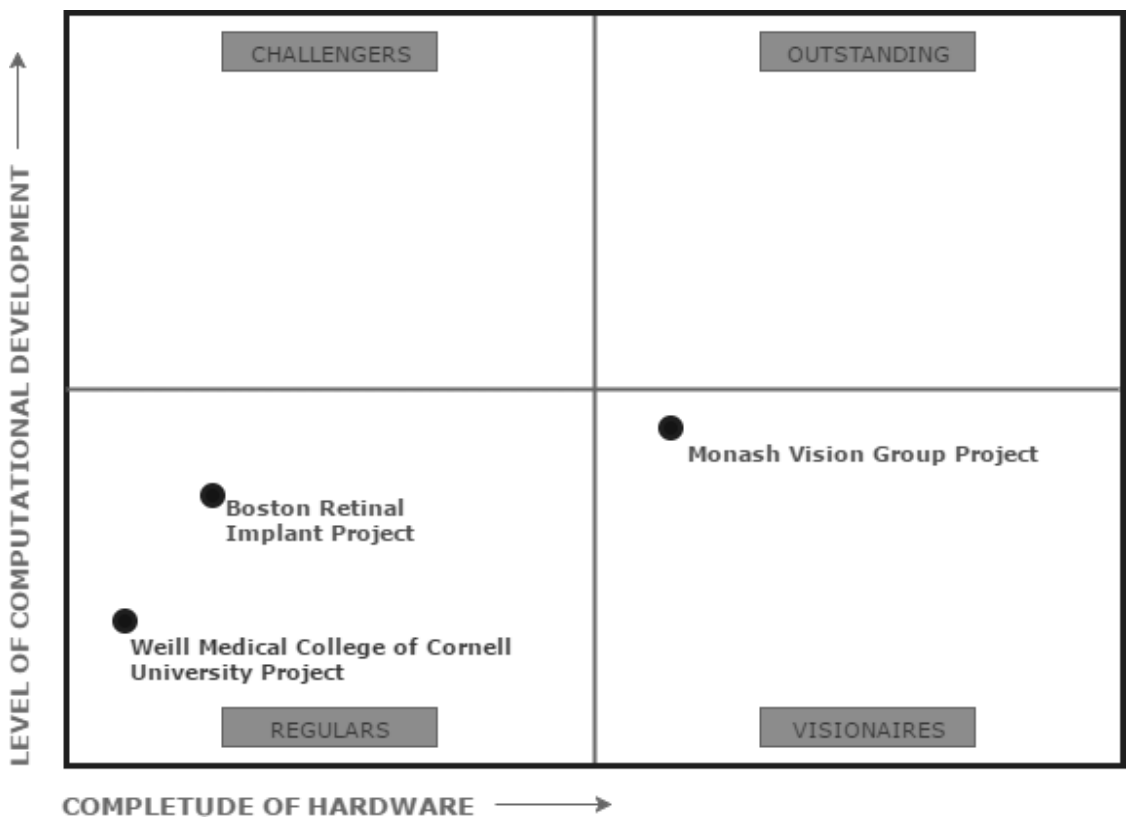

Font: Authors.